\documentclass[preprint2,longabstract]{aastex}

\usepackage{enumerate}

\newcommand{\lya}{Ly$\alpha$}
\newcommand{\halpha}{H$\alpha$}
\newcommand{\hbeta}{H$\beta$}
\newcommand{\paalpha}{Pa$\alpha$}
\newcommand{\brgamma}{Br$\gamma$}

\newcommand{\flya}{$F_{\mathrm {Ly}\alpha}$}
\newcommand{\wlya}{$W_{\mathrm {Ly}\alpha}$}

\newcommand{\ebv}{$E_{B-V}$}
\newcommand{\ebvs}{$E_{B-V,\star}$}
\newcommand{\ebvg}{$E_{B-V,\mathrm{IS}}$}
\newcommand{\hi}{H{\sc i}}
\newcommand{\hii}{H{\sc ii}}

\newcommand{\ctn}{$CTN$}
\newcommand{\chisq}{$\chi^2$}

\newcommand{\lamp}{$\lambda_{\mathrm p}$}
\newcommand{\rwid}{$W_{\mathrm r}$}

\newcommand{\nneb}{$n_\mathrm{neb}$}
\newcommand{\nfs}{$n_\mathrm{fs}$}

\slugcomment{Accepted by the Astronomical Journal}

\shorttitle{Continuum subtracting HST \lya\ images}
\shortauthors{Matthew Hayes et al.}

\begin{document}

\title{Continuum subtracting Lyman-alpha images: 
	Low redshift studies using the Solar Blind Channel of HST/ACS }

\author{Matthew Hayes}
\email{matthew.hayes@unige.ch}
\affil{Geneva Observatory, University of Geneva, 51 chemin des Maillettes,
 	1290 Sauverny, Switzerland} 
\author{G{\"o}ran {\"O}stlin\altaffilmark{1}}
\email{ostlin@astro.su.se}
\affil{Stockholm Observatory, AlbaNova University Centre, SE-106\,91 Stockholm, Sweden} 
\author{J. Miguel Mas-Hesse\altaffilmark{2}}
\affil{Centro de Astrobiolog{\'a} (CSIC-INTA), E28850 Torrejon de Ardoz, Madrid, Spain}
\email{mm@laeff.inta.es}
\author{Daniel Kunth}
\affil{Institut d'Astrophysique de Paris, Paris (IAP), 98 bis boulevard Arago, 75014 Paris, France}
\email{kunth@iap.fr}

\altaffiltext{1}{Oscar Klein Centre for Cosmoparticle physics, Department of Astronomy,
  Stockholm University, Sweden.} 
\altaffiltext{2}{Laboratorio de Astrof{\'i}sica Espacial y F{\'i}sica Fundamental
(LAEFF-INTA), POB 78, Vva. Canada, Spain}

\begin{abstract}
We are undertaking an imaging study of local star-forming galaxies in the
Lyman-alpha (\lya) emission line using the {\em Solar Blind Channel (SBC)} of 
the {\em Advanced Camera for Surveys} onboard the {\em Hubble Space Telescope}. 
Observations have been obtained in \lya\ and H-alpha (\halpha) and six
line-free continuum filters between $\sim 1500$\AA\ and the $I-$band. 
In a previous article
\citep{Hayes05}
we demonstrated that the production of \lya\ line-only images (i.e. continuum
subtraction) in the {\em SBC}-only data-set 
is non-trivial and that supporting data is a requirement. 
We here develop various methods of continuum subtraction and assess their 
relative performance using a variety of spectral energy distributions (SED) as
input. 
We conclude that simple assumptions about the behavior of the ultraviolet
continuum consistently lead to results that are wildly erroneous, and
determine that a spectral fitting approach is essential.
Moreover, fitting of a single component stellar or stellar+nebular spectrum is
not always sufficient for realistic template SEDs and, in order to successfully 
recover the input observables, care
must be taken to control the contribution of nebular gas and any underlying
stellar population. 
Independent measurements of the metallicity must first be obtained, while
details of the initial mass function play only a small role.
We identify the need to bin together pixels in our data to obtain 
signal--to--noise $(S/N)$ ratios of around 10 in each band before processing. 
At $S/N=10$ we are able to recover \lya\ fluxes 
accurate to within around 30\% for \lya\ lines with intrinsic equivalent width
(\wlya) 
of 10\AA.
This accuracy improves to $\lesssim 10$\% for \wlya=100\AA.
We describe the method of image processing applied to the observations presented in 
\"Ostlin et al. (2009) and the associated data-release.
We also present simulations for an observing strategy for an alternative 
low-redshift \lya\ imaging campaign using {\em ACS/SBC} using adjacent
combinations of long-pass filters to target slightly higher redshift. 
\end{abstract}


\section{Introduction}

The Lyman-alpha emission line (\lya) is a powerful and frequently exploited 
observational signature through which the galaxy population can be probed 
at high-redshift ($z$).
In principle \lya\ can be used to probe the ionization fraction during the final
stages of re-ionization 
\citep{Malhotra04,Dijkstra07}, 
cosmic star-formation rates
\citep{Hu98,Kudritzki00,Ajiki03}, 
large scale structure
\citep{Venemans02,Ouchi05}, 
and to identify potential host of population {\sc iii} star formation
\citep{Malhotra02,Nagao07}. 
Ultimately, exactly how the \lya-emitting (or non-emitting) high-$z$ galaxy 
population relates to the ultraviolet-selected Lyman break galaxy (LBG) 
population is uncertain, and studies of the high-$z$ population of \lya\ emitters
(LAEs) are interesting in their own right. 

In all surveys for which the ultimate science goal is more fundamental than the
observed population itself, it is vital to understand what biases, be they
observational or astrophysical, may affect the inferred properties 
and how they manifest themselves. 
This point is especially consequential for high-$z$ \lya-selected studies 
where typical detections are faint and galaxies often go undetected in the 
continuum. 
\lya\ is a resonant line and its formation is strongly affected by a complex
radiative transport.
On a physical level the regulation and transport of \lya\ is known to be
affected by dust, the topology and ionization of the ISM, \hi\ kinematics,
and viewing geometry.
Such insights have been gleaned empirically from spectroscopic observations of 
small samples of local galaxies 
\citep{Giavalisco96,Kunth98,Mas-Hesse03}, 
flanked by a sophisticated theoretical and computational attack on 
the problem of \lya\ escape physics
\citep{Neufeld90,Ahn04,Hansen06,Verhamme06,Laursen07}.

Since \lya\ is a resonance line and can be expected to be substantially
spatially decoupled from UV continuum radiation and other nebular lines, 
the picture yielded by UV-targeted spectroscopy is limited. 
Thus the imaging approach becomes an invaluable complement to the previous 
{\em IUE} and {\em HST} spectroscopic studies. 
\lya\ imaging at $z\approx 0$ is technically possible with {\em HST} using the 
{\em Wide Field and Planetary Camera (WFPC2)} although the instrumental
throughput at \lya\ would make it very inefficient.
A much more economical approach would be to use the UV-optimized channels of the
{\em Space Telescope Imaging Spectrograph (STIS)} or 
{\em Advanced Camera for Surveys (ACS)} which both offer higher system
efficiency and better defined and non red-leaking filters. 
In {\em HST} cycle 11 we began an imaging study to examine a handful
of local star-forming galaxies in \lya\ using {\em HST/ACS}.
First results from this were presented in 
\cite{Kunth03},
although technical uncertainties about how to subtract the continuum prevented a
deep analysis and 
\cite{Hayes05} 
demonstrated the need for additional off-line observations to aid in the
continuum subtraction.
This additional data has since been obtained for the remainder of the sample and 
the entire data-set has now been processed. 
\lya\ line-only images are being released to the community 
(\"Ostlin et al. 2009). 
However, since the process of subtracting the continuum in this study is far
from trivial and future imaging studies depend upon the 
methodology, this process was deemed to be worthy of an article in its own right. 
We here perform a series of tests of synthetic \lya\ imaging observations of 
low-$z$ targets using configurations available on {\em HST} and compare
different approaches to subtracting the continuum. 

The article is organized as follows: 
in Sect.~\ref{sect:history} we explain the complications involved with \lya\
observations and why conventional continuum subtraction techniques are not
applicable to \lya\ with the current instrumentation; 
in Sect.~\ref{sect:methtest} we describe the methodology and theoretical 
tests; 
in Sect.~\ref{sect:resdisc} we present the results and discuss their
significance; 
in Sect.~\ref{sect:future} we present simulations for a study targeting
slightly higher redshift  and discuss some possible augmentation to the method;
and 
in Sect.~\ref{sect:conc} we make some concluding remarks.

\section{Local Lyman-alpha imaging} \label{sect:history} 

\subsection{Observing strategies}\label{sect:obsstrat}

\begin{figure}
\epsscale{.99} 
\plotone{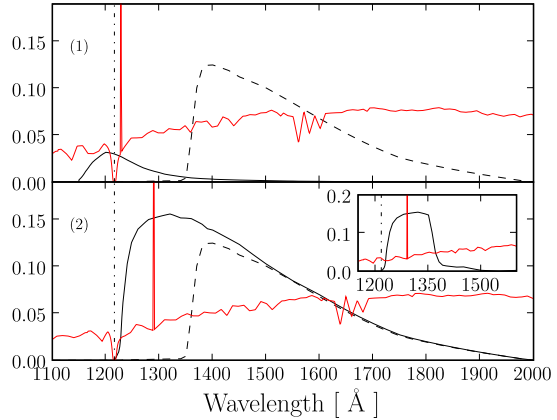}
\caption{{\em ACS/SBC} filter combinations for \lya\ imaging. 
Solid lines represent the filter that transmits \lya\ and dashed lines that for
the continuum.
Vertical dot-dashed lines are at $\lambda = 1216$\AA, the wavelength of both
Milky Way \lya\ absorption and the geocoronal emission line. 
The red line shows a synthetic spectrum with artificially added \lya. 
For illustration, a Milky Way absorption feature has also been added to the 
templates ($\log ( n_\mathrm{HI} ) = 20.5$~cm$^{-2}$).
{\em Upper}: 
Configuration 1: {\em F122M+F140LP} for $z\lesssim 0.03$. 
Suitable for $z\approx 0$ although both geocoronal emission and Galactic
absorption fall in the on-line bandpass and must be accounted for. 
The spectrum is redshifted to 0.01.
{\em Lower}: 
Configuration 2: {\em F125LP+F140LP} for $0.03\lesssim z \lesssim 0.1$. 
Geocoronal \lya\ is not transmitted and on-line system
throughput is greatly improved over the {\em upper} example.
The spectrum is redshifted to 0.06. 
}
\label{fig:filters}
\end{figure}

The {\em Solar Blind Channel} ({\em SBC}) of {\em HST/ACS}
 offers the {\em F122M} filter through which to observe restframe \lya\ at 
$0 < z \lesssim 0.035$.
This filter peaks at 1216\AA\ and has a rectangular width (\rwid) of 128.4\AA\ 
although the pivotal wavelength (\lamp) at 1273.7\AA\ is rather offset from the peak 
due to the red wing. 
Furthermore, with \rwid/\lamp$\approx 0.1$, this filter cannot be considered 
narrow compared to optical narrowband filters. 
A number of long-pass filters are available through which to sample the
continuum, the most appropriate for $z\approx 0$ being {\em F140LP} 
(\lamp=1527\AA; \rwid=252.95\AA). 
This {\em F122M+F140LP} configuration is that used by our imaging campaign 
and is denoted configuration 1 in this article. 
Due to the very sharp cut-on on the blue side, and the red side being defined
by the degrading sensitivity of the detector with wavelength, all the 
long-pass filters exhibit near identical shapes in the red wing.
Adjacent long-pass filter combinations could therefore be appropriate for 
\lya\ imaging of slightly more distant targets
(e.g. {\em F125LP+F140LP} for $0.028 \lesssim z \lesssim 0.1$) and this
combination is also examined here, denoted as configuration 2.
Fig.~\ref{fig:filters} shows the bandpasses for both configurations.

\subsection{Continuum subtraction methods}\label{sect:contsub_method}

The art of continuum subtraction hinges entirely upon scaling of the
stellar continuum sampled by an off-line filter to the bandpass of the
filter that isolates the line. 
In the forthcoming discussion we make use of a quantity known as the {\em
Continuum Throughput Normalization (CTN)} factor, first defined in 
\cite{Hayes05}. 
This is the dimensionless quantity that scales the raw count-rate in a continuum 
filter to that
expected in the on-line filter, accounting for the filter transmission profiles,
the instrument sensitivity, and the shape of the continuum. 
That is, 
\begin{equation}
\mathrm{Ly}\alpha = \mathrm{online} - CTN \times \mathrm{offline}
\end{equation}
\ctn\ can be computed with varying degrees of complexity.
For known or assumed continuum slopes it can be computed simply from the inverse
sensitivity bandpass characteristics (PHOTFLAM) of the on- and off-line filters. 
For any continuum spectrum, \ctn\ can be computed by convolving the spectrum
with the instrument throughput profiles of the filters, integrating to estimate 
the count-rate, and calculating the ratio.
This permits features such as spectral shape and absorption or emission lines 
to be accounted for in the process.
In essence \ctn\ is simply a color with a non-standard normalization
derived from instrument sensitivities. 

There are a number of possible issues that may impede the continuum subtraction of 
\lya.
Firstly, possible strong stellar \lya\ absorption may cancel
some or all of the nebular emission.
This is perhaps unlikely to be a strong affect where the starburst is very young
and dominated by O-stars where deep absorption features have yet to develop. 
The effect may become very significant when \lya\ is scattered and may be superimposed 
upon older stellar populations with lower effective temperature. 
Indeed, absorption of stellar continuum by B-stars was initially proposed as
an explanation for the weakness or absence of \lya\ emission in nearby
starbursts and for the early failure to detect \lya\ emitters at high-$z$
\citep{Valls-Gabaud93}.
Secondly, the \lya\ feature may be P\,Cygni with some or all of the emission 
cancelled by absorption, especially directly in front of the brightest clusters.
Spectroscopic measurements, if of sufficient resolution, can isolate the 
emission segment of the P\,Cygni profile and obtain the emitted flux,
although this is not possible with imaging. 
If the absorption segment falls within the on-line filter as it will in our case, 
the local minima in \lya\ flux from P\,Cygni absorption on top of clusters can never 
be corrected for. 
Notably however, this caveat also applies to imaging surveys at high-$z$.
Ultimately what is called emission is a matter of definition; whether it be
defined as the flux in the emission segment of the profile, or the flux
integrated over both the absorption and emission segments.
Either way, imaging can only assess the latter. 
Finally, and especially for configuration 1, photometry will be directly affected by 
Milky Way \lya\ absorption and geocoronal \lya\ emission.
Both of these effects can be relatively easily accounted for since they are not
expected to vary strongly across the small angles subtended by the {\em SBC} 
field--of--view.
Geocoronal emission can be removed by background subtraction, and Milky Way
\lya\ absorption computed from the H{\sc i} column density measured along the
line--of--sight to the target 
\citep{Dickey90,Kalberla05}
and incorporated into the calculation of \ctn.

\subsubsection{Purely observational
		methods}\label{sect:purelyobservationalmethods}

Typically, for optical emission lines  in ground-based
observations, continuum is sampled using a filter positioned as
close as possible to, but not transmitting, the line in question (or other
contaminating spectral features).
For example, an off-line \halpha\ filter 100\AA\ redward of the line corresponds
to $\Delta\lambda / \lambda \sim 0.015$.
Were an intrinsically power-law continuum with $\beta=-2$ assumed to be flat
($\beta=0$), this would result in an error in the line-center continuum estimate of around 
3\%\footnote{Continuum is assumed to take the form of a power-law in
$\lambda$ of the form $f_\lambda \propto \lambda^\beta$}.
Typically, this would be considered `good enough'. 

The same is not true for the FUV filter set available on {\em HST/SBC}. 
Due to the limited choice of filters available, for the {\em
F122M+F140LP} the on- and off-line combination, 
the pivot wavelengths are separated by $\Delta\lambda / \lambda \sim 0.2$ which
would result in errors of $\gtrsim 30$\% if the same misguided assumption of the
continuum slope were made.
Errors in line-center continuum estimation of this magnitude could translate
into severe errors in continuum subtraction of \lya\ since the on-line bandpass
is so broad: if the line does not dominate, then the continuum subtracted pixel
could quite feasibly get the wrong sign (emission could be seen as absorption
and vice versa). 
For a dust-free starburst the slope of the FUV continuum is largely constant 
$\beta \sim -2.6$ over the first few 10s Myr, but flattens
quite rapidly ($\beta$ increases) thereafter, particularly at higher 
metallicities 
\citep{Leitherer99}. 
Since the UV slope is also a strong function of the dust reddening (\ebv), it 
is not possible to reliably predict $\beta$ globally, let alone how it varies 
on $\sim 10$~pc scales (i.e. pixel--to--pixel, at {\em SBC} sampling).
One possible alternative would be to sample the blue and red sides of the 
on-line filter but there are no filters available for such an observation and
spectroscopic observations with the {\em STIS} have shown that the continuum on
the blue side of \lya\ is frequently unpredictable, contaminated by internal or
Galactic absorption features. 
Flux at \lya\ due to continuum processes must be estimated from observations 
on the red-side of \lya\ only. 

The next step would be to take an additional off-line observation redwards of the
off-line filter (e.g. {\em ACS/SBC/F150LP} or {\em F165LP}, {\em WFPC2/F218W}
{\em ACS/HRC/F220W}, or {\em WFC3/UVIS/F225W}) and extrapolate a power-law 
continuum to the on-line filter.
This way \ctn\ would be computed for a given spectral slope by convolving
the throughput profiles with the measured power-law, resulting in a different
\ctn\ in each pixel. 
However, this method was demonstrated to result in significant errors in the
continuum flux estimation when relatively moderate amounts of dust are present
\citep{Hayes05,Hayes06} 
since typical dust reddening modifies the FUV continuum in such a way that 
it becomes inconsistent with the power-law approximation. 
Furthermore, such a method provides no estimate of stellar \lya\ absorption
which is significant in all but very hottest stars.

Naively, these points could be argued away:
only the very youngest star-forming regions produce enough ionizing photons to 
generate significant \lya\ 
\citep{Charlot93}
over which times $\beta$ is essentially constant. 
High resolution imaging studies have however demonstrated that in the ISM of
active starbursts, particularly surrounding massive young clusters, the nebular
emission can be clearly displaced from the ionizing sources 
\citep[e.g.][]{Maiz-Apellaniz04}. 
This is most likely due to stellar winds and supernova feedback clearing bubbles
in the ISM and is clearly visible in some cases in our sample, well exemplified
by ESO\,338-IG04
(\"Ostlin et al. 2009).
The case is further complicated for \lya\ by its resonant nature.
As a result of multiple scatterings in H{\sc i}, \lya\ photons diffuse away from
their production sites and are likely emitted from a site of last scattering
that is not coincident with the nebulae where they were produced
(\citealt{Hayes05,Hayes07}; \"Ostlin et al. 2009). 
\objectname[]{ESO\,338-IG04}
\citep{Hayes05}
shows a large diffuse emission component which dominates the
total \lya\ luminosity. 
High resolution imaging frequently reveals that starbursts are composed of
numerous compact young knots and clusters that dominate the luminosity but do
not account for a large fraction of the surface area
\citep{Meurer95}, 
particularly at UV wavelengths. 
It is therefore rather likely that the continuum that needs to be subtracted 
will not be representative of a region young enough to produce \lya, which
can emerge superimposed upon regions that appear too dusty for \lya\
to be transmitted 
\citep{Hayes07}. 

\begin{figure}
\epsscale{.9} 
\plotone{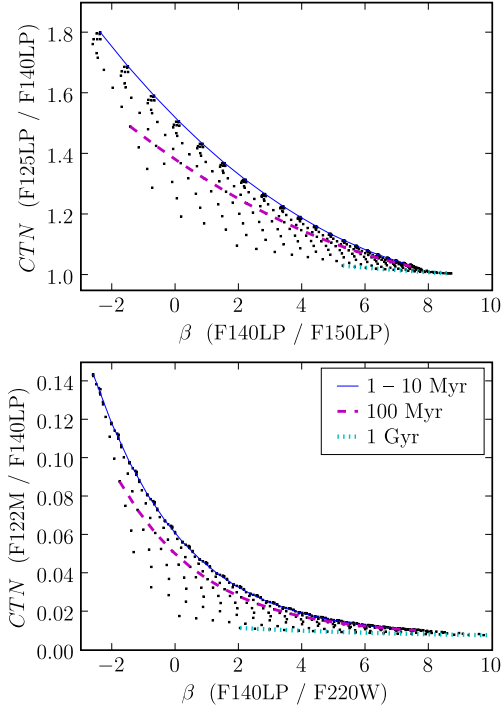}
\caption{{\em CTN vs.} $\beta$ for various ages (1\,Myr to 1\,Gyr) and \ebv\ in
the range 0 -- 1 for the law of 
\cite{Calzetti94}. 
Lines link points of the same age (i.e. follow tracks of dust reddening). 
{\em Upper}: The configuration used for our current {\em HST} campaign. 
\ctn\ is generated for {\em F122M}/{\em F140LP} with $\beta$
measured between {\em F140LP} and {\em F220W}. 
\ctn\ is clearly not a monotonic function of $\beta$ and 
degenerate values of \ctn\ can be derived for any measured $\beta$, especially 
for bluer colors (e.g. $\beta \sim -1$).
{\em Lower}: Same as upper but for the configuration 2. 
\ctn\ is generated for {\em F125LP/F140LP} with $\beta$
measured between {\em F140LP} and {\em F150LP}. 
The degeneracy still exists although has been quite significantly improved. 
} 
\label{fig:ctnbeta}
\end{figure}

Figure~\ref{fig:ctnbeta} demonstrates inadequacy of assuming a one-to-one
relationship between \ctn\ and $\beta$. 
Here the two quantities are computed for various ages and \ebv\ by
convolving {\em Starburst99} spectra \citep{Leitherer99,Vazquez05}
with the filter throughput profiles.
For this demonstration the 
\cite{Calzetti94} 
attenuation law is adopted which is frequently used to represent dust
extinction in starbursts and LBGs.
The first example shows our own observing strategy using {\em SBC/F122M} for \lya\
on-line, {\em SBC/F140LP} for off-line, measuring $\beta$ between {\em F140LP} and
{\em HRC/F220W}. 
The second example shows how, by adopting the {\em F125LP}, {\em F140LP} and 
{\em F150LP} configuration, the degeneracy in \ctn\ {\em vs.} $\beta$ is significantly 
reduced.
However, with uncertainties still greater than 40\% at certain $\beta$, it is
still not sufficient for the broad on-line filter. 
In general the extinction laws of the Milky Way and Magellanic clouds are 
found to be steeper than the 
\cite{Calzetti94}
law at given \ebv\ over the wavelength domain covered by our filters
and the effect is found to be more severe.

\subsubsection{Methods beyond pure observation}

\ctn\ is the only important quantity to know for the continuum subtraction
but unfortunately, as demonstrated in 
Section~\ref{sect:purelyobservationalmethods}, 
the quantity cannot easily be estimated without some knowledge of the continuum.
A solution is compute \ctn\ from spectral synthesis models using
supplementary data to constrain the model.
Thus one must have sufficient 
information at hand to build the right stellar population; stellar age and
star-formation history, \ebv, metallicity, and the initial mass function (IMF)
may all have a significant effect.

\begin{figure*}
\epsscale{1} 
\plotone{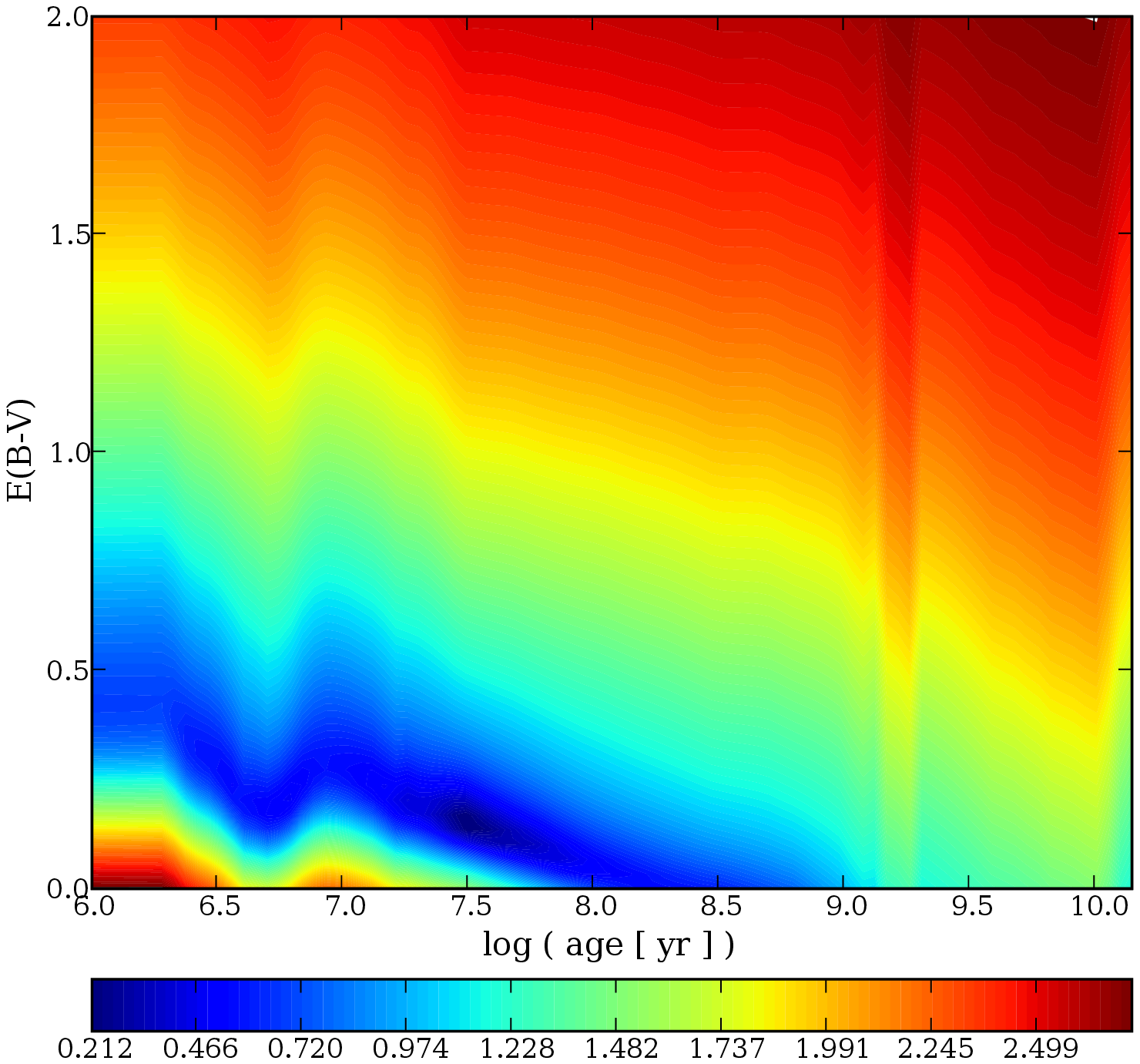} 
\epsscale{1} 
\plotone{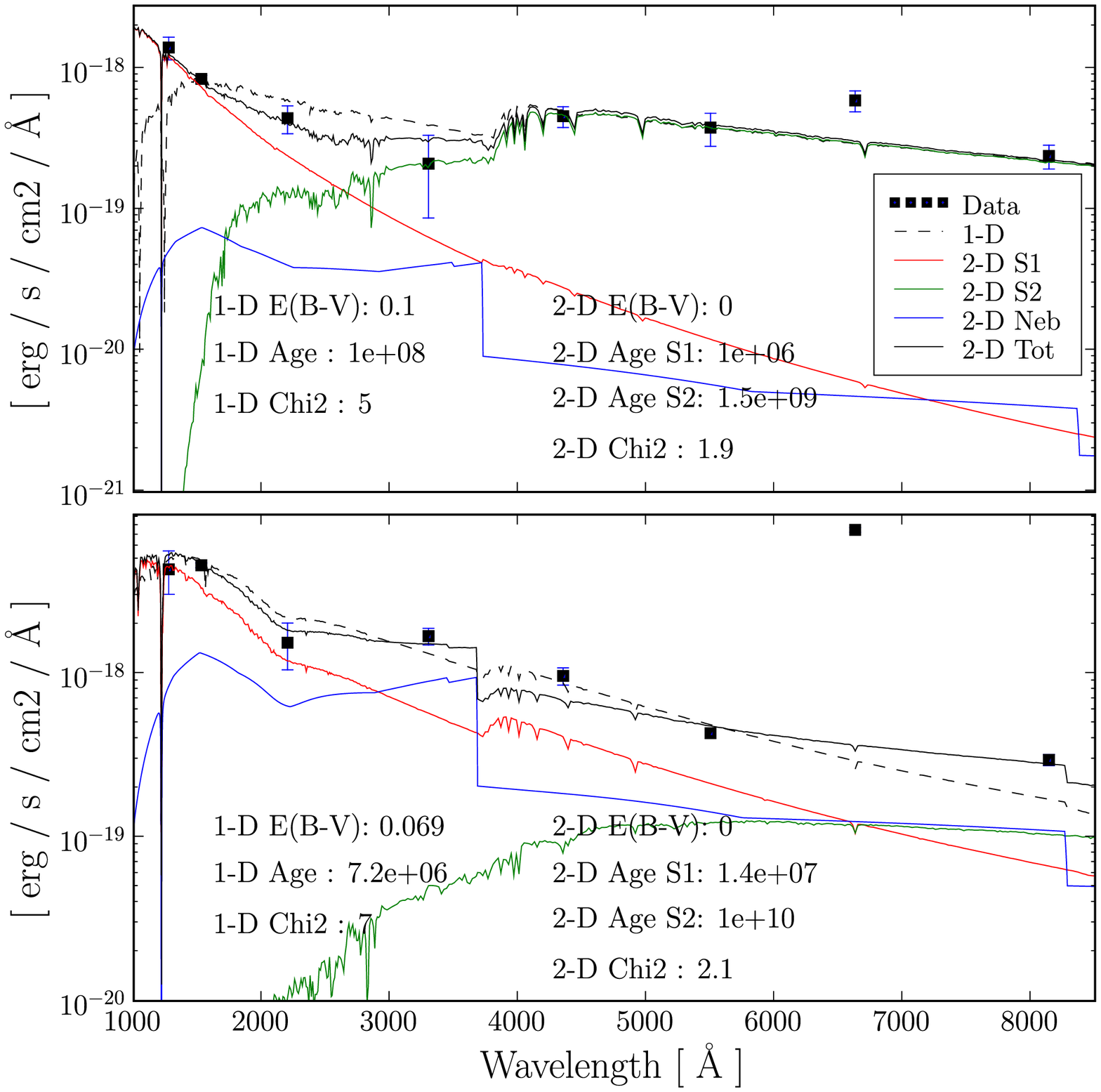} 
\caption{{\em Left}: $\log(\chi^2)$ map for a one component stellar fit of
age and \ebv.
A degeneracy is clearly seen, running diagonally and negatively
in age--\ebv\ space although note the logarithmic scaling.
Data for this plot is taken from a diffuse \lya\ emission region in the vicinity
of Knot B of Haro\,11.
{\em Upper right}: Example of a multi-component SED fitting to real data from Haro\,11
\citep{Hayes07}.
Blue shows the nebular gas continuum spectrum, the level
of which is determined from the \halpha\ observation (note the data point at
\halpha\ is lies high above the fit, since this observation contains both line
and continuum), green shows the underlying field-star population of age around
1~Gyr
that dominates in the $I-$band, red shows the current starburst with an age of
1~Myr. Black shows the total combined spectrum. 
Dashed black line shows a single component spectral fit.
{\em Lower right:} Same as {\em upper} but extracted from a well resolved 
\halpha\ shell in ESO\,338-IG04
(\"Ostlin et al. in prep.).
The extremely high \halpha\ equivalent width is apparent since the stellar
continuum is rather weak.
The nebular continuum level is higher than that of the stars in the $U$ and $I$ bands. 
The data-points show the $U-B$ color is negative and the Balmer jump is significant in 
the total spectrum. The nebular component still does not contribute near \lya\ but 
demonstrates why control over the nebular continuum spectrum is necessary. 
}
\label{fig:ageebv}
\end{figure*}

In 
\cite{Hayes05} 
we demonstrated that we could reliably continuum subtract \lya,
pixel--by--pixel,
in our {\em HST} observations (configuration 1) through multi-color
spectral modeling
using a number of additional broadband {\em HST} observations. 
In that study four line-free observations were used, together with the {\em
Starburst99} spectral evolutionary models.
In age--\ebv\ space we showed that we could find non-degenerate solutions for
\ctn\ by sampling $\beta$ (sensitive to both age and dust) and the 
Balmer/4000\AA\ break (sensitive primarily to the age of the stellar 
population). 
Grids of multiple color indices and \ctn\ were computed from the models for a 
range of age and \ebv, enabling us to look up the most appropriate value of 
\ctn\ in each pixel. 
We also demonstrated that several other parameters (metallicity,
IMF, etc.) had only a small impact upon the reliability of the method. 
If \chisq\ is defined as 
\begin{equation}
\chi^2 = \sum_i \left[ C\cdot m_i(t, E_{B-V}) - d_i \right] ^2 \cdot W_i 
\label{eq:chisq1d}
\end{equation}
where $m_i(t,E_{B-V})$ represents the $i$th model data point on the SED for given 
age $(t)$ and \ebv, $C$ the 
model normalization factor, $d_i$ flux measured in the $i$th filter, and 
$W_i$ the weight for the corresponding filter, 
typically the inverse variance.
The value of $C$ that provides the best fit is determined from 
\begin{equation}
C = \frac{\displaystyle \sum_i m_i d_i W_i}{\displaystyle \sum_i m_i^2 W_i} 
\label{eq:solve1d}
\end{equation}
where $C$ and \chisq\ are computed for each possible model spectrum, selecting
the best models from the minimum value of \chisq. 
Figure~\ref{fig:ageebv} shows an example \chisq\ map in $t$--\ebv\ space.

Naturally, if \ctn\ is sensitive to the Balmer/4000\AA\ break, then a number of
contaminants may affect the \ctn\ determination, depending upon how well the
stellar and nebular emission regions are resolved. 
At the earliest times when the nebular emission is strong, the Balmer edge may
contribute by bluening the restframe $U-B$ color. 
On the other hand, at later times the 4000\AA\ break arises from metal line blanketing 
in late type stars, reddening $U-B$.
Thus resolved nebular gas and old stellar populations may both affect 
the SED in the optical but not the FUV; not affecting \ctn\ itself but its
determination should optical SED data-points be relied upon.
Since the relative contribution from stellar populations and nebular gas is
unknown, and in high-resolution imaging the regions may be clearly resolved, 
it may become necessary to account for all these populations in our estimation of 
the true spectrum, and therefore \ctn.

Control over the nebular gas contribution may be obtained through additional
observations that directly trace the ionized gas. 
For given temperature in an optically thin nebula, the 
emission coefficients ($\alpha^{eff}$) may be computed for any allowed
recombination line, and 
continuous emission coefficient ($\gamma_\nu$) computed for given wavelength. 
Thus an appropriately strong and well-behaved line (e.g. \halpha) can be used
to estimate the nebular continuum contribution in each filter, and subtract it
from each data-point for a given \ebv\ (again, the filters have been selected in 
order to avoid the strongest nebular emission lines).
\ctn\ may then be computed using the SED-fitting method to
obtain the stellar-only SED (minimizing $\chi^2$ using 
Equation~\ref{eq:chisq1d}) and reconstruction the composite spectrum.

The current starburst may be superimposed upon any underlying stellar
population.
Since this is likely to show a significant 4000\AA\ break but little FUV
contribution, it may affect the reliability of \ctn\ determination that relies
upon age fitting. 
Thus it may be necessary to treat more than one stellar population per
pixel.
Old stellar populations may dominate the integrated light at red wavelengths
and such an observation may facilitate the decomposition of starburst and
underlying stellar components.
Equation~\ref{eq:chisq1d} can be modified to include two populations, $a$ and
$b$, each with differing normalization as 
\begin{equation}
\chi^2 = \sum_i \left[ C_a \cdot m_{a,i} + C_b \cdot m_{b,i} - d_i \right] ^2 \cdot W_i 
\label{eq:chisq2d}
\end{equation}
where $m_{a,i}$ and $m_{b,i}$ represent the model SED data-points, with $C_a$ and
$C_b$ the normalization factors for each population. 
For a given set of models $m_{a}$ and $m_{b}$, an analytic solution exists for
the values of $C_a$ and $C_b$ that provide the best-fit, as: 
\begin{eqnarray}
C_a &=& \frac{S_5 S_2 - S_3 S_4}{S_2^2 - S_1 S_4} \nonumber \\
C_b &=& \frac{S_3 - C_a S_1}{S_2}
\label{eq:solve2d}
\end{eqnarray}
where
\begin{eqnarray}
S_1 &=& \sum_i m_{a,i}^2 W_i   \nonumber \\
S_2 &=& \sum_i m_{a,i} m_{b,i} W_i   \nonumber \\
S_3 &=& \sum_i m_{a,i} d_i W_i   \nonumber \\
S_4 &=& \sum_i m_{b,i}^2 W_i   \nonumber \\
S_5 &=& \sum_i m_{b,i} d_i W_i   
\label{eq:Ses}
\end{eqnarray}

With the contributions from the two stellar populations measured or fit, the
total SED can be reconstructed by summing the SED in each component.
Thus \ctn\ is determined from the reconstructed model spectrum. 

The fact that stellar ages may be estimated provides the first-order solution to
another potential problem: that of the underlying stellar \lya\ absorption feature. 
For O stars, and depending on the wind properties, \lya\ can be expected in
absorption, filled in P\,Cygni, or emission (net equivalent widths of
a few \AA\ are expected, \citealt{Klein78}). 
However, due to the deep Galactic \hi\ column to even the nearest O stars, the
intrinsic \lya\ profile is completely absorbed and it is likely these
predictions can never be empirically tested.
Some B stars however are near enough to be observed at \lya\ and NLTE atmosphere models
appear to be able to match the observed profile (\lya\ absorption equivalent widths of 
a few tens of \AA).
Since no observational tests are available, it is not deemed meaningful to pursue
this issue further, although it is worth pointing out that the stellar feature
is accounted for, within the limits of current understanding.

In the following Sections we assess the relative power of the various methods
described above in recovering \lya\ fluxes and equivalent widths.

\section{Methodology and Tests}\label{sect:methtest}

The aim of this study it to test how reliably we can recover \lya\ observables
from our imaging observations.
Ideally we would like to test the various methods of continuum subtraction
against real observational data for which the true \lya\ fluxes and equivalent
widths are known (i.e. spectra). 
Unfortunately there are no real object spectra that span the wavelength
range between \lya\ and 9000\AA\ with sufficient spectral resolution, taken in
consistent apertures for us to test the various methods against real data, and 
we are forced to rely only upon synthetic input sources.

In this study we present a number of computational tests to assess the
performance of various methods. 
To this end we generate a set of template spectra using various
combinations of {\em Starburst99} models 
\citep{Leitherer99,Vazquez05}, 
for various stellar populations (or combinations thereof), \ebv, and modify 
them by adding \lya\ lines of chosen equivalent width. 
We then convolve the template spectra with the {\em HST} bandpasses 
to generate synthetic SED data-points and feed these into our SED-fitting
and continuum subtraction software. 
Thus we have full knowledge of the intrinsic spectrum, and we are able to compare 
our output results directly to the known input values (\ctn, \flya, \wlya), and 
test the ability of our fitting methods to recover ages and reddening in the stellar 
population(s).

We simulate noise by randomizing each SED point for a given error
and, adopting a Monte-Carlo approach, we compute various statistics of the
recovered distributions. 
For the real {\em HST} data we employ adaptive binning techniques to obtain
a minimum threshold $S/N$, but the choice of this value requires testing. 
Therefore by performing these simulations we are 
able to test the results obtained against input $S/N$, and 
directly test the optimum threshold.  

The method of generating the `real' input spectra is outlined in
Section~\ref{sect:sedgen}.
The various methods of subtracting the continuum are described in
Section~\ref{sect:contsub}, and the actual quantitative tests are described in
Section~\ref{sect:tests}.

\subsection{Input SED generation}\label{sect:sedgen}

We have previously discussed the fact that SED is likely to 
be the sum of the contributions from a current episode of star formation
(starburst; {\tt sb}), an underlying component of field stars ({\tt fs}),
and emission from nebular gas ({\tt neb}). 
All template input spectra are generated from a combination of these three
components, scaled to a given normalization relative to the starburst at
4500\AA\ ($\sim B-$band); these normalizations are assigned the \nfs\ and \nneb\
for the field stars and nebular components, respectively.
The templates thereby consist of two stellar components of 
variable ages, and a single gas component, all contributing different
fractions of the $B-$band luminosity. 
For simplicity, the two stellar components are generated from the same
metallicity and initial mass function. 
The \halpha\ luminosity is computed from the nebular gas continuum flux density and 
applied to the gas spectrum before addition. 
To complete the restframe spectrum, it is reddened using a given extinction law 
and \ebv.
Finally the spectrum is redshifted.  
Further information on the generation of input spectra can be found in 
Section~\ref{sect:tests}
and 
Table~\ref{tab:parameters}.

For each template spectrum, we generate a set of `observed' SED data-points by
convolving the spectra with the {\em HST} filter profiles. 
For configuration 1 (our current imaging campaign)
the complete filter list is: {\em SBC/F122M} (\lya\ on-line), {\em SBC/F140LP}
($\sim 1500$\AA\ continuum), {\em HRC/F220W} ($\sim 2200$\AA\ continuum),
{\em HRC/F330W} ($\sim U-$band), {\em WFC/F435W} ($\sim B-$band), 
{\em WFC/F550M} (medium $V-$band, serving as line-free continuum filter near
\halpha), {\em WFC/FR656N} (linear ramp narrow-band filter centered upon
restframe \halpha), and {\em HRC/F814W} ($\sim I-$band). 
For configuration 2  the filter set
replaces {\em F122M} with {\em F125LP} while {\em F140LP} remains the
FUV continuum filter as shown in Figure~\ref{fig:filters}. 
For the optical component of this configuration we 
sample the UV/optical continuum using {\em WFPC2/F336W}, 
{\em F439W}, and {\em F814W}, using the fixed narrow-band filter {\em F673N} to
observe redshifted \halpha. 

\subsection{Continuum subtraction methods}\label{sect:contsub}

For each SED we apply various methods of estimating and subtracting the
continuum and examine how well they return the known restframe quantities 
that were input. 
We identify five possible methods of estimating \ctn\ and subtracting the
continuum at \lya: 
\begin{enumerate}[I]

\item{Assuming the slope of the continuum between the off-line and 
on-line filters to be  flat in $f_\lambda$. 
I.e. \ctn\ is the ratio of the {\em HST} PHOTFLAM values for each filter. }

\item{Assuming the continuum takes the form of a power-law in $f_\lambda$
($\propto \lambda^\beta$) and extrapolating the slope as measured between
the NUV filter at $\sim 2200$~\AA\ and the off-line FUV filter ({\em F140LP}).}

\item{Fitting a single {\em Starburst99} spectrum (combined stellar+nebular 
components)
to the continuum SED points (i.e. excluding the \halpha\ observation) using
Equations~\ref{eq:chisq1d} and ~\ref{eq:solve1d}.
Then computing \ctn\ from the best-fitting (age and \ebv) spectrum.  }

\item{Continuum-subtracting \halpha\ and using the \halpha\ flux to reconstruct 
the SED due to nebular continuum processes. 
Then fitting a single stellar-only {\em Starburst99} spectrum (again fitting age and 
\ebv) to the real SED data-points with the nebular contribution
subtracted (i.e. real data-points -- nebular data-points) using the same method as
method {\sc iii}.  }

\item{Same as method {\sc iv} but fitting 2 stellar components to the
nebular-subtracted SED using Equations~\ref{eq:chisq2d},\ref{eq:solve2d}, 
and \ref{eq:Ses}.
In this case, the fitting is performed over all ages with age allowed to
vary in {\it both} stellar populations.  
}

\end{enumerate}

\subsection{The tests}\label{sect:tests}

The most important returned values are, of course, \flya\ and \wlya, although 
from methods where SED fitting is employed, age(s), normalization factor(s), 
and \ebv\ are also returned. 
Monitoring these returned values in addition to those relating to \lya\ permits
a deeper examination of the performance of the SED fitting.

We begin by defining a fiducial template spectrum, denoted {\tt fid}. 
This is generated from what is
thought to be a typical starburst, capable of producing \lya\ with moderate
extinction and metallicity and a Salpeter IMF. 
For the starburst, a single stellar population of age of 5~Myr is selected 
since ionizing photons are a requirement for the production of \lya.   
According to the {\em Starburst99} template spectra, the nebular gas
contribution at 4500\AA\ (\nneb, defined relative to the stellar spectrum of and
unresolved point-source) is $\sim 0.05$ at this age, and this value provides the 
fiducial \nneb. 
For the fiducial model we add a moderate field-star population with age
(arbitrarily selected)
5~Gyr, scaled to give a contribution of 50\% the starburst luminosity at 
4500\AA\ (\nfs=0.5). 
 The fiducial reddening for the composite spectrum was selected to be 
\ebv$=0.2$ using the SMC law. 
The value of \ebv\ is an approximate midpoint in the measured extinction for
most of our galaxies (see 
\citealt{Atek08}). 
The choice of law is motivated by the fact that the SMC law appears to provide the best
fits to the resolved spectra of blue compact and irregular galaxies 
\citep[e.g.][]{Mas-Hesse_Kunth99}.
Since \wlya\ can essentially take any value between damped absorption and 
super-recombination values, we add \lya\ lines with a range of equivalent 
widths: 
-50\AA\ to represent the deep absorption seen in some local objects 
(e.g. {\sc i}\,Zw\,18);
0\AA, corresponding to no \lya\ feature;
10\AA\ to approximate the global values measured in some of low-$z$ objects	
\citep[e.g.][]{Giavalisco96}; 
and 100\AA\ corresponding to high-$z$ \lya-bright galaxies or diffuse emission
regions in resolved objects at $z\approx 0$. 
We define a number of modifications to all of the ingredient parameters in
the input spectrum which are listed in  Table~\ref{tab:parameters}. 

Firstly, all tests are performed at infinite $S/N$ to test the reliability of
the code by insuring all the input parameters are returned.

\begin{deluxetable}{lllll}
\tabletypesize{\scriptsize}
\tablecolumns{5}
\tablewidth{0pt}
\tablecaption{Parameters for SED generation}
\tablehead{ \colhead{Parameter} & \colhead{Unit} & \colhead{Fiducial value} &
	\colhead{Non-fiducial / Range} & \colhead{Code} } 

\startdata
$W_{\mathrm{Ly}\alpha}$ & \AA\      & 10                & -50, 0, and 100  & 
      {\tt [ fid\_w10; fid\_w-50; fid\_w0; fid\_w100 ] }\\
\ebv\                   & mag       & 0.2               & 0.0 and 0.5      & 
      {\tt [ ebv0\_w...; ebv05\_w...] } \\
Reddening law           & ---       & SMC               & \cite{Calzetti94} & 
      {\tt [ cal\_w...] } \\
Starburst age           & Myr       & 5                 & 20 and 100         & 
      {\tt [ agesb20\_w...; agesb100\_w...] } \\
Field star age          & Myr       & 5000              & 200                & 
      {\tt [ agefs200\_w... ]}  \\
\nfs~$^1$    & ---       & 0.5               & 0 and 5             & 
      {\tt [ nfs0\_w...; nfs5\_w...] } \\
\nneb~$^2$   & ---       & 0.05               & 0 and 1       & 
      {\tt [ nneb0\_w...; nneb1\_w...] } \\
IMF $\alpha$            & ---       & $-2.35$           & $-1.85$ and $-2.85$ & 
      {\tt [ imf185\_w...; imf285\_w...] } \\
Metallicity             & $Z$       & 0.08               & 0.001 and 0.040          & 
      {\tt [ metsub\_w...; metsuper\_w...] } \\
Stellar atmosphere~$^3$ & ---       & Atm5              & Atm3 and Atm4       & 
      {\tt [ atm3\_w...; atm4\_w...] } \\
Redshift                & ---       & 0.01              & ---                 &
--- \\

\enddata
\tablecomments{Model parameters in the third column constitute the 
{\tt fid\_w10} fiducial template spectrum case, with the non-fiducial values 
for \wlya\ constituting standard cases of {\tt fid\_w-50}, {\tt fid\_w0}, and 
{\tt fid\_w100}.\\
$^1$~defined relative to Starburst population at 4500\AA\ (i.e. $\sim$
		relative $B-$band luminosity). \\
$^2$~defined relative to Starburst population at 4500\AA. 
0.05 corresponds to typical default unresolved nebular fraction at 5Myr from 
{\em Starburst99}.  0 and 1 correspond therefore to zero
nebular contribution and a `boost' by a factor of $\sim 20$. \\
$^3$ Designated {\em Starburst99} codes. 
Atm3 is Lejeune atmospheres for stars with plane-parallel atmospheres and
Schmutz atmospheres for stars with strong winds.
Atm4 is like Atm3 but with Hillier atmospheres for stars with strong winds.
Atm5 is like Atm4 with Pauldrach models for O stars.}
\label{tab:parameters}
\end{deluxetable}

Observational $S/N$ per resolution element varies significantly with position 
as a result of natural morphological variation in surface brightness, and
band-to-band as morphology differs with wavelength.
In our {\em ACS} data-set we make use of adaptive binning 
to bin together pixels until a threshold $S/N$ has been met, conserving surface 
brightness in each conglomerate bin (``spaxel").
Without investigation it is not possible to know what $S/N$ is required
in order for the various techniques to become optimal; either converging on the
input values or to some systematically offset values.
By varying $S/N$ in our simulations, we are able to test this directly. 
To each data-point in the input SED (see 
Section~\ref{sect:sedgen}), we assign an error, based upon the $S/N$ 
we want to test. 
Using this error we then regenerate 1000 SEDs using Gaussian-variates and apply
all our continuum subtraction methodologies to each one. 
Performing such Monte-Carlo simulations with increasing $S/N$ allows us to
investigate the optimum $S/N$ required to return mean values consistent with 
the those input, or find convergence of the mean to systematically offset 
observables. 
For each of the model galaxy SEDs defined in Table~\ref{tab:parameters}, we run
1000 Monte-Carlo iterations at each $S/N$ with $S/N$ varying between 1 and 50.
Important statistics of \flya\ and \wlya; the mean and median, standard 
deviation, and skewness of the derived distribution are retained. 

Firstly we set $S/N$ to be the same in all bandpasses;
if the worst quality band is chosen to generate the binning pattern, $S/N$ in the
corresponding bins in other bands should have $S/N$ exceeding the threshold.
However, due to differing morphologies, this is unlikely to be the case in
general, and some bins, in some images, may fall short of the desired $S/N$
threshold. 
We assess the impact of this to the overall fitting by systematically
dropping $S/N$ to zero in individual bands for the fiducial SED. 
This allows us to examine the robustness of returned \lya-related quantities 
in the case where lower-quality data have been obtained in certain bandpasses.
Similar tests are then performed dropping $S/N$ to zero in adjacent pairs of filters
simultaneously. 

To test the reliability of the methods for a wide array of model galaxies we
augment the fiducial SED, one parameter at a time, always running the
SED-fitting and continuum subtraction code using the standard {\em Starburst99}
parameters and SMC reddening law.
Age and reddening, of course, are always the free parameters in the fitting 
(both age components are fit for method {\sc v}).
This way we can assess the reliability of our results at different stellar ages,
\ebv, and with different contributions from the underlying stellar population
and nebular gas. 
The same is also true when we test metallicity, IMF, reddening law, enabling us
to test the impact of failing to select the correct values for the fitting. 
The effect of metallicity is found to be have sufficient impact that
additional tests are performed, see 
Section~\ref{sect:resultsmetallicity}.
The parameter space covered is listed in Table~\ref{tab:parameters}. 
Parameter dependency tests are performed for both observational configurations
with a detailed discussion of the results for configuration 1, including the
effects of poor $S/N$ in certain filters, presented in Sect~\ref{sect:resdisc}. 
A summary of the results for configuration 2 is presented in 
Sect~\ref{sect:future}.

\section{Results and discussion}\label{sect:resdisc}

\subsection{The fiducial model}

In this subsection we first present and discuss the results obtained for the 
fiducial SED with the same $S/N$ in all bandpasses in Sect~\ref{sect:goodsn},
followed by results where single and adjacent pairs of filters have
$S/N=0$ in Sect~\ref{sect:poorsn}.

\begin{figure*}
\epsscale{1}  
\plotone{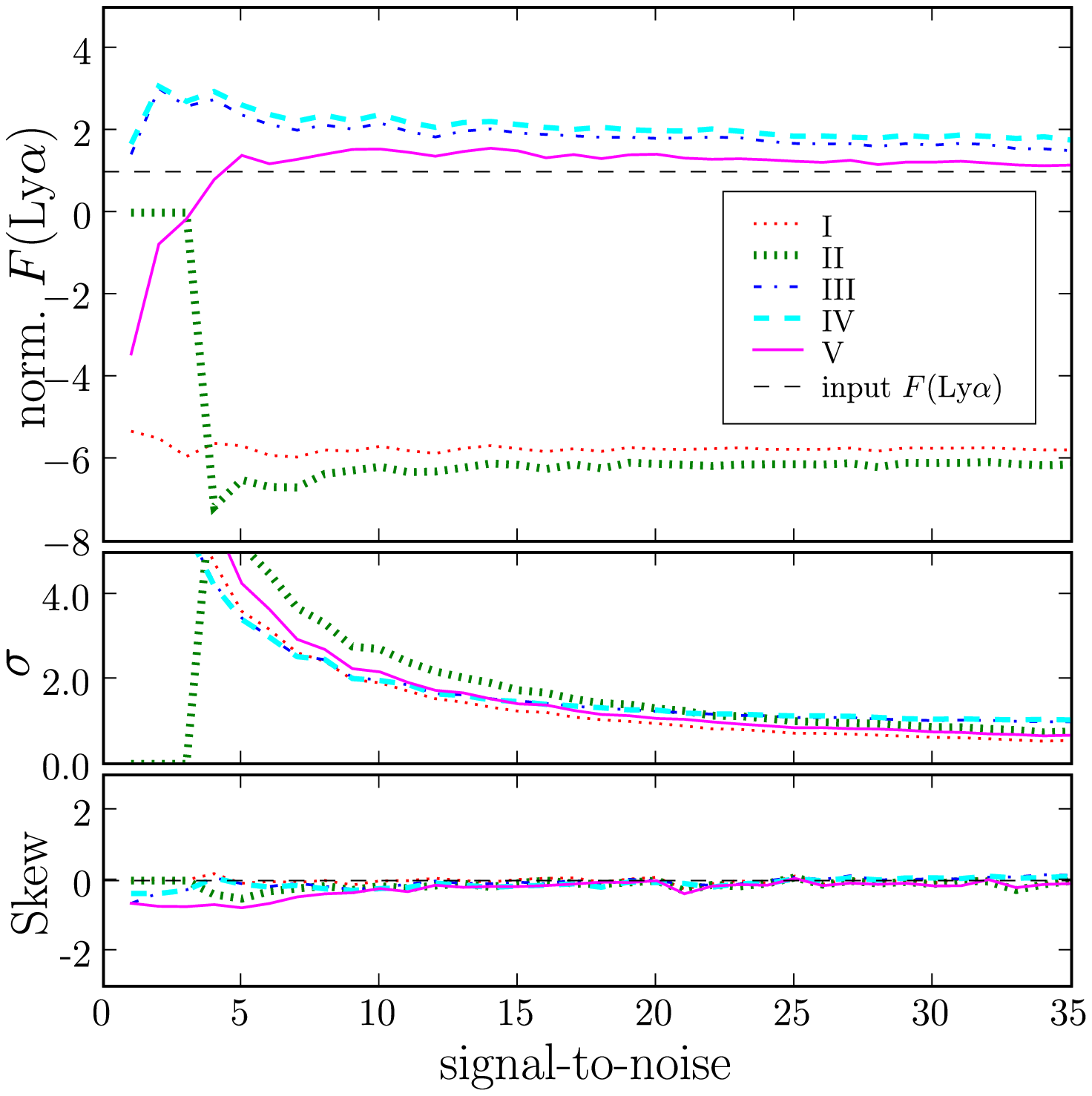}
\plotone{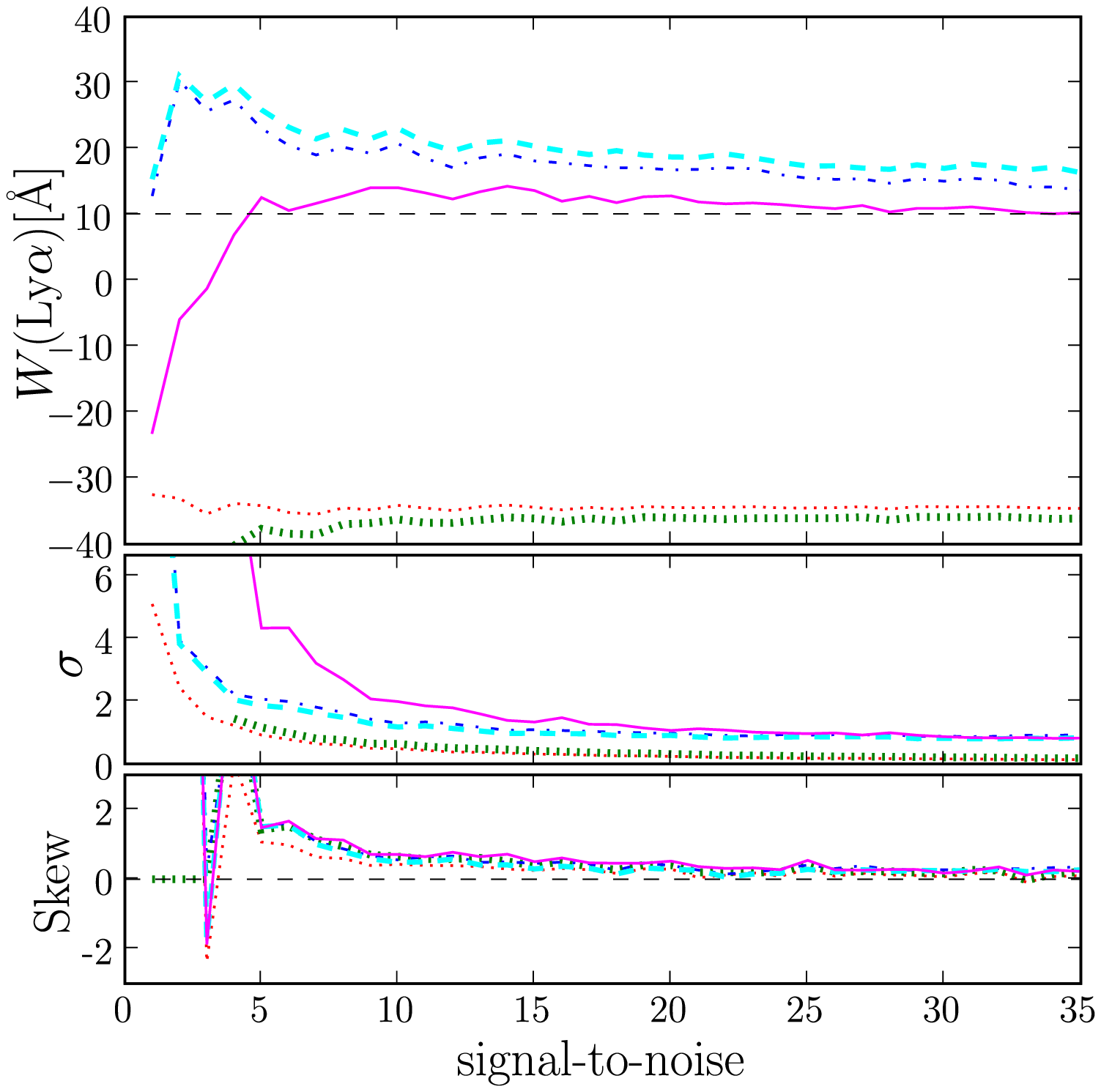}
\caption{Statistics of the returned distribution from 1000 Monte-Carlo 
iterations on the fiducial setup with \wlya\ of 10\AA. 
{\em Top} panels show the mean, {\em center} the
$1\sigma$ standard deviation, and {\em lower} the skewness.
{\em Left}: Continuum subtracted \lya\ flux in normalized units (i.e. returned
mean divided by input flux). 
Standard deviation is also given in normalized units. 
{\em Right}: \lya\ equivalent width. 
Colors, line styles, and weights show the various methods of estimating the continuum and 
are labeled in the legend. 
}
\label{fig:std_w10_s2n}
\end{figure*}

\subsubsection{Equal $S/N$ in all bands}\label{sect:goodsn}

Figure~\ref{fig:std_w10_s2n} shows the returned statistics for the fiducial
model as a function of $S/N$. 
Statistics shown include the mean ({\em upper}), standard deviation 
({\em center}), and skewness ({\em lower}).
Firstly, these plots demonstrate that simple assumptions about the
continuum slope (either $\beta=0$ (method {\sc i}) or $\beta$ extrapolation
(method {\sc ii})) both result
in an overestimate of \ctn.
Therefore the continuum is over-subtracted and a 10\AA\ emission line is seen 
as an absorption feature with \wlya$\sim -40$\AA.
This is the result of: (a) the large offset in $\lambda$ between the off- and 
on-line filters; (b) the fact that the $\sim 100$\AA\ wide on-line bandpass is 
far from line-dominated; and (c) the modest value of \ebv$=0.2$ is sufficient to 
reduce the total flux in {\em F122M} below that of {\em F140LP}. 
It is clear that a good understanding of the continuum is essential, not only
between the filters but across the on-line bandpass itself, and that
spectral modeling of some level of sophistication is a requirement. 
It should be noted that at $S/N=10$ in the individual bandpasses, $S/N$ in
the continuum subtracted \lya\ distribution 
is only around 0.5, although obviously this
improves with increasing \wlya\ as the line starts to dominate. 
For the {\tt fid\_w100} case, $S/N=4$ is seen in the returned \lya\ flux
distribution for $S/N=10$ in all filters. 
Of the three SED-fitting methods, technique {\sc v} slightly out-performs {\sc iii} 
and {\sc iv}, thanks to its inclusion of treatment of the underlying
stellar population and nebular gas, even though they are only minor
contributors to the FUV flux for the {\tt fid} SED. 
It should also be noted that all the continuum subtraction methods show a 
positive skew in the \wlya\ distribution, even at $S/N > 30$. 
This is due to fact that $W$ is a ratio, and that the inverse of a Gaussian
distribution always shows positive skewness -- this should be present in all
equivalent width estimates, irrespective of data-set or observational
methodology 
\citep[ for a discussion see][]{Dawson04,Hu04,Hayes06}.

An estimate of the age of the stellar population also provides, in part, the
solution to another potential problem, that of the unknown underlying stellar 
absorption at \lya.
While still poorly tested, \lya\ features in the models still provide the best 
estimate of the underlying absorption available and are included in computed
values of \ctn\ and therefore the estimate of the continuum flux.

\subsubsection{Reduced filter sets or poor $S/N$}\label{sect:poorsn}

\begin{figure}
\epsscale{1.1} 
\plotone{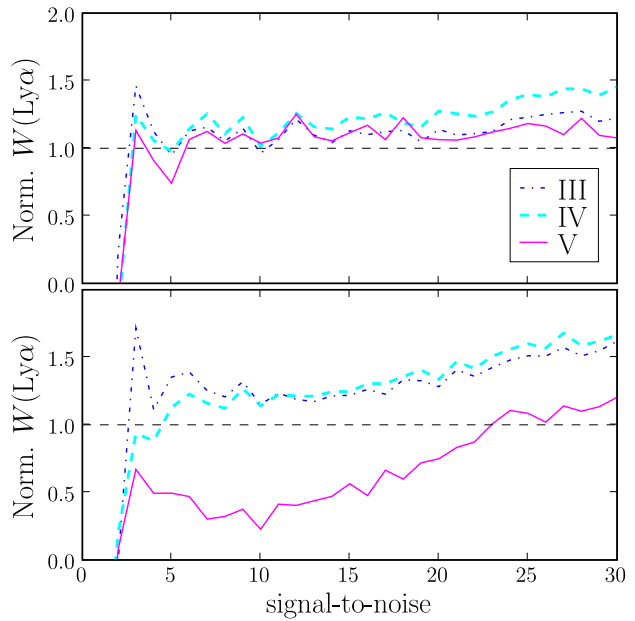}
\caption{Average values of \wlya\ returned by the SED fitting software for the
fiducial model galaxy with \wlya=10\AA\ when datapoints are removed from the
fitting procedure.
$S/N$ is the same in all the remaining bandpasses. 
\wlya\ is now normalized by its value when all filters are included, where
convergence was previously found after $S/N\sim 5$ 
(see Figure~\ref{fig:std_w10_s2n}). 
The {\em upper} panel shows the removal of the {\em F220W} filter. 
The removal of both {\em F220W} and {\em F330W} is shown below.
}
\label{fig:std_w10_rat_f220w_f330w}
\end{figure}

The upper panel of Figure~\ref{fig:std_w10_rat_f220w_f330w} shows the effect of
removing a single filter from the fitting routine for the {\tt fid\_w10} SED. 
Naturally, only the three continuum subtraction methods that employ SED fitting
are shown. 
We now show the mean \wlya\ obtained with the filter removed, normalized by that
obtained with all filters included, with the same $S/N$ in all included 
bandpasses. 
This normalized distribution has clearly been shown to converge in flux at 
$S/N\approx 5$ (Figure~\ref{fig:std_w10_s2n}). 
The results are noisier than those shown previously for two reasons: firstly
because the fits are intrinsically noisier due to one fewer data-point being
present, and secondly because of added noise from the normalization. 
The example shown illustrates the removal of the {\em F220W} filter, although
results are largely indistinguishable when other single filters are removed. 
This is because we have a maximum of 5 model parameters to fit 
and in the case
where a single filter is removed, we still have
sufficient data-points remaining to avoid a degeneracy.
Provided we obtain the minimum $S/N$ threshold of 10 as shown in
Section~\ref{sect:goodsn} in 5 of our 6 continuum bandpasses, we can feel safe
about the recovery of \lya\ fluxes. 

The situation changes, however, when two filters are removed from the fit as can
be seen in the lower panel of Figure~\ref{fig:std_w10_rat_f220w_f330w}.
This plot shows the same as the upper panel when the two UV filters 
({\em F220W} and {\em F330W}) are assigned $S/N=0$. 
The two single stellar component fitting methods ({\sc iii} and {\sc iv}) now
consistently overestimate \wlya\ by around 20\% at $S/N \sim 10$ which 
actually becomes worse as $S/N$ increases in the other filters.
This is due to more consistent selection of the wrong stellar parameters by 
the code and the example has been selected to show poorly recovered 
observables; the loss of both UV filters is the most detrimental. 
Without either {\em F220W} or {\em F330W}, we have no sampling of the FUV 
continuum slope or the Balmer/4000\AA\ break, essentially resulting in the 
recovery of any ages and \ebv. 
Clearly maintaining $S/N\sim5$ sampling of the Balmer break is preferable 
to obtaining extremely high $S/N$ observations in other bandpasses; as 
discussed in 
\cite{Wiklind08}
the Balmer break is instrumental in resolving the degeneracy between age and 
reddening.
We cannot conclude that methods {\sc iii} and {\sc iv} out-perform {\sc v} in
this case but we can be quite certain that we need to reach the threshold $S/N$
in 5 of our 6 continuum bandpasses in order for any of the SED-fitting methods
to yield robust results. Four filters is not deemed to be sufficient for method
{\sc v}.

\subsection{Parameter dependencies}

We now assess the impact of modifying the various parameters that go into the
construction of a the composite galaxy spectra. 
The fiducial parameters are always used internally for the fitting (third column
in Table~\ref{tab:parameters}), fitting only ages and \ebv. 

\begin{figure*}
\epsscale{1} 
\plotone{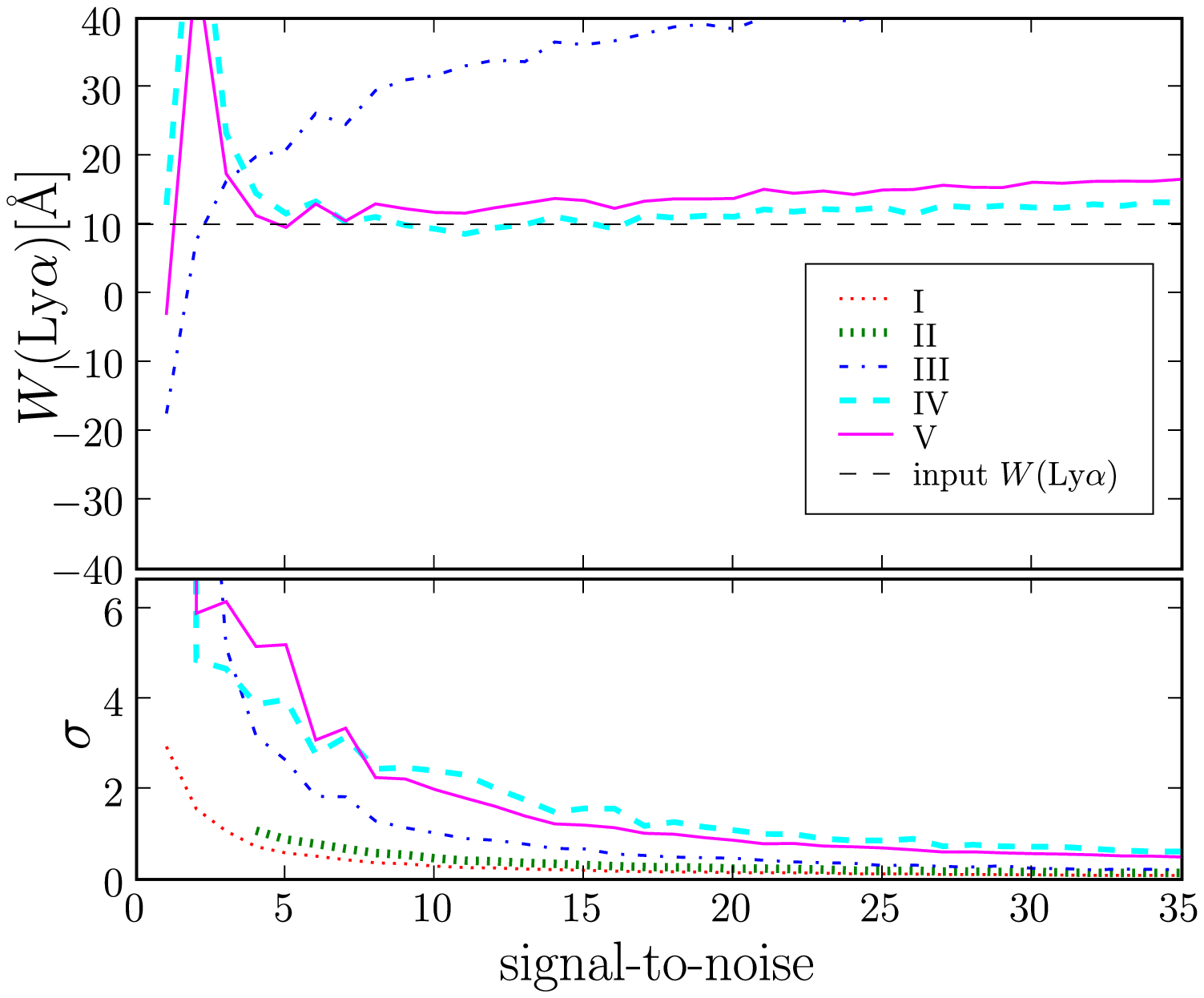}
\plotone{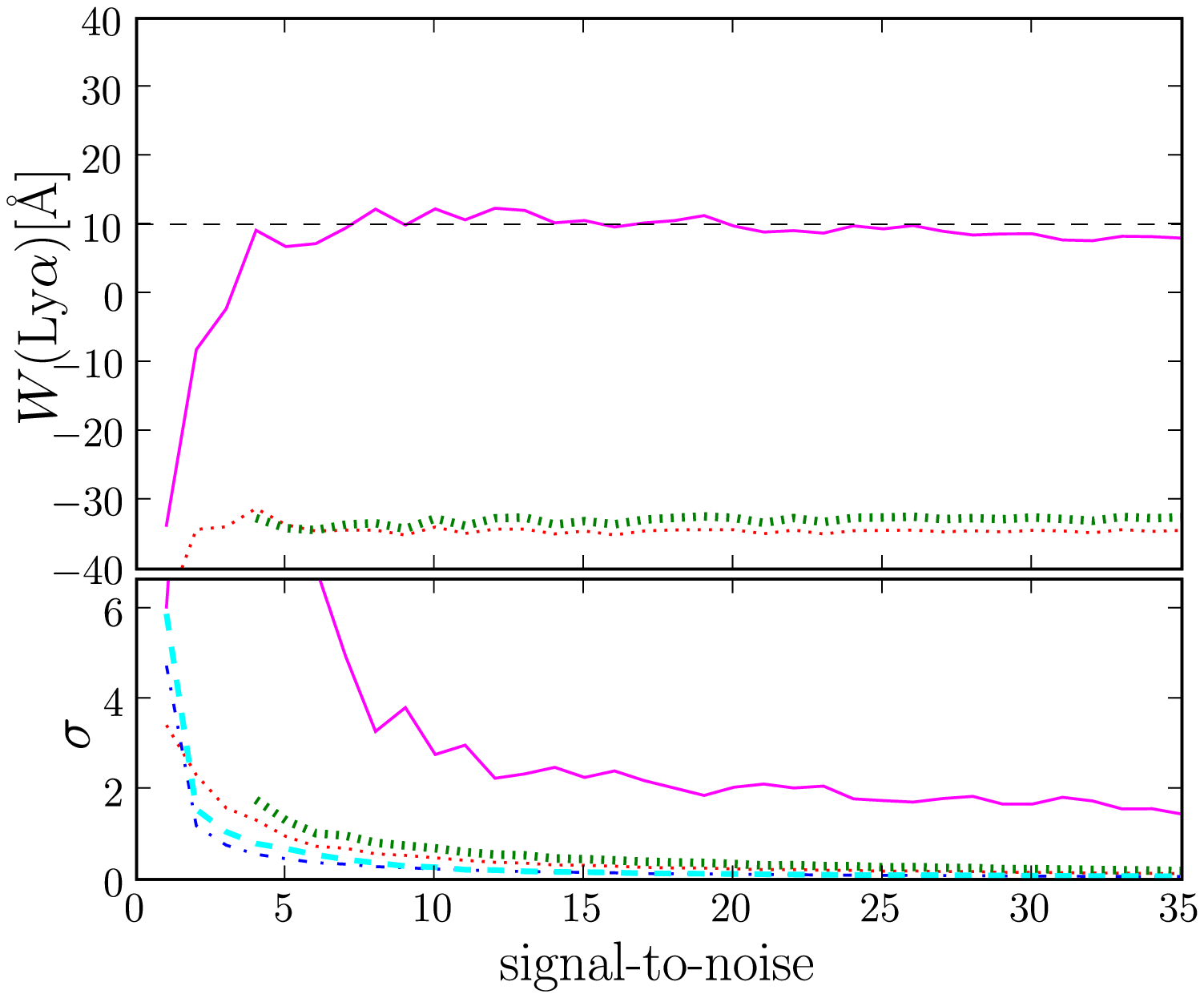}
\caption{As in the right panel of Figure~\ref{fig:std_w10_s2n} but with
different input parameters. 
{\em Left}: {\tt nneb1\_w10} in which the starburst and nebular gas component
contribute equally at 4500\AA.  
Methods {\sc i} and {\sc ii} both converge significantly off the lower end of
the ordinate axis. 
{\em Right}: {\tt nfs5\_w10} in which the field star population outshines the
starburst at 4500\AA\ by a factor of 5.  
Methods {\sc iii} and {\sc iv} both converge significantly off the ordinate axis
at equivalent widths of almost 200\AA.  }
\label{fig:std_w10_nneb_nfs}
\end{figure*}

\subsubsection{Stellar ages and dust reddening}

Table~\ref{tab:parameters} shows the various parameter modifications made to the
{\tt fid} galaxy template for starburst age, field star age, and \ebv.
For the SED-fitting methods ({\sc iii}, {\sc iv}, and {\sc v}) results are found
not to be be significantly discrepant from those presented for the fiducial
case.
In fact, aside from details in the noise, the flux and equivalent width plots are
indistinguishable from those in Figure~\ref{fig:std_w10_s2n}.
This is not the case for methods {\sc i} and {\sc ii} which vary wildly with
starburst age and \ebv.
Naturally the behavior of the modeling methods is to be expected since age and 
\ebv\ are our primary fitting parameters and the effect of dust can be well 
controlled in the event that the chosen extinction law well describes the intrinsic 
deviation from the dust-free starburst. 

Exchanging the SMC extinction law for the attenuation law of 
\cite{Calzetti94}
is more detrimental:
these two curves are not near equivalent over the wavelength domain we are 
sampling. 
For the {\tt cal\_w10} case, techniques {\sc iii} and {\sc iv} result in
convergent \wlya\ estimates of 75 and 112\AA\ respectively while method {\sc v}
converges at \wlya=20\AA\ by $S/N=10$ (still an overestimate by a factor of 2). 
For strong emission ({\tt cal\_w100}), method {\sc v} overestimates 
\wlya\ by just 10\% at $S/N=10$ while {\sc iii} and {\sc iv} still consistently
fail by a factor of 2. 
However, clearly an accurate model of the internal extinction law in the target 
galaxy is a key parameter in the method. 

Several solutions to this exist. 
Firstly, the reddening law itself could be incorporated as a free 
parameter, only increasing computation time by a factor of a few.
Including individual extinction laws (e.g. SMC, LMC, or Galactic curves) in the SED 
fitting is physically rather poorly motivated and 
a better alternative may be to base the choice of law on observation. 
For example FUV spectroscopy of the objects could indicate the presence or
absence of a 2175\AA\ graphite feature to motivate the choice of curves (see
\citealt{Puget_Leger89} 
and the discussion in 
\citealt{Mas-Hesse_Kunth99}).

\subsubsection{Nebular gas and underlying stellar population}

Figure~\ref{fig:std_w10_nneb_nfs} shows the average returned \wlya\ as a
function of $S/N$ when the nebular gas
component contributes equally with the starburst at 4500\AA\ 
({\tt nneb1\_w10}; {\em left}), 
and when the underlying population contributes 5 times that of the starburst
({\tt nfs5\_w10}; {\em right}). 
The left plot demonstrates how a single component fit ({\sc iii}) breaks down in
the {\tt nneb1\_w10} model when
the nebular gas component is dominating the SED (i.e. regions where wind-blown 
\hii\ shells or filamentary structure are resolved).
Method {\sc iii} converges with a returned \wlya\ of around 40\AA\ in this case,
and a similar arithmetic ($+30$\AA) overestimate for all the input equivalent 
widths. 
This demonstrates the need for an independent measure of the nebular gas
spectrum as methods {\sc iv} and {\sc v} well recover \wlya.

The right plot shows the necessity to also control the underlying stellar
population when fitting stellar ages. 
The ordinate axis does not show methods {\sc iii} or {\sc iv} which both
converge at \wlya$\sim 200$\AA.
This corresponds to an overestimate of an order of magnitude, with  
overestimates seen for all input \wlya.
Notably, \wlya\ converges at around 400\AA\ in the {\tt nfs5\_w100} case and
almost +100\AA\ in the {\tt nfs5\_w-50} case using methods {\sc iii} and 
{\sc iv}.

\subsubsection{Metallicity} \label{sect:resultsmetallicity}

\begin{figure}
\epsscale{1.1} 
\plotone{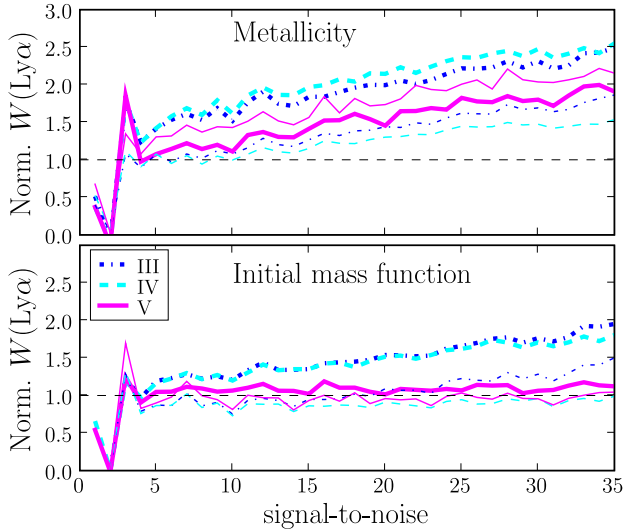}
\caption{As in Figure~\ref{fig:std_w10_rat_f220w_f330w} for different
metallicities ({\em upper}) and initial mass functions  ({\em lower}).
Input \wlya\ is 10\AA. 
SED-fitting methods only are shown and methods are labeled. \wlya\ is normalized
by the resulting values obtained from the fiducial model, output values of which
have previously been shown to converge at around $S/N\sim 5$. 
For metallicity, the thick lines represent sub-solar and the thin lines
super-solar metallicity. 
Regarding the IMF, the thick lines show the flatter {\tt imf185\_w10} model while
the thin lines show the steeper {\tt imf285\_w10} model.
}
\label{fig:std_w10_metimf}
\end{figure}

Two additional metallicities were initially tested: the minimum sub-solar value 
($Z=0.001$; {\tt metsub})
and maximum super-solar value 
($Z=0.040$; {\tt metsuper}). 
The {\em upper} panel of Figure~\ref{fig:std_w10_metimf} shows the returned
values of \wlya\ for the models with modified metallicities {\tt metsub\_w10}
and {\tt metsuper\_w10}, again normalized by the returned values for the 
{\tt fid\_w10} model as described for Figure~\ref{fig:std_w10_rat_f220w_f330w}. 
Interestingly, while one represents a decrease in metallicity and one an
increase, both modifications have a similar impact on the 
resulting observables: the continuum flux in the on-line filter is consistently
underestimated. 
The underestimate is around 5\% in the both cases but, for the weak emission 
cases this is enough to cause overestimates in \wlya\ of 10\AA\ 
(a factor of 2) using method {\sc v} and worse for {\sc iii} and {\sc iv}. 
The effect is less noticeable in the cases of stronger \lya\ emission where
the underestimate of the on-line continuum flux linearizes and converges to
overestimates proportional to the error on the continuum.
In a similar manner to when filters were removed from the fit
(Sect.~\ref{sect:poorsn}), increasing $S/N$
means that unrepresentative fits are found more consistently, resulting in less
spread in the output values than at low $S/N$ where the spread is the result of
noisy fits.
In the case of metallicity, poor recovery of stellar age is found to be the 
culprit behind deviant \lya\ observables due to the impact of metallicity on the
4000\AA\ break. 

The apparent strong effect of unconstrained metallicity appears to be a
concern and has prompted further tests. 
In the second round of testing, all available metallicities were examined; $Z=[0.001, 0.004,
0.008, 0.020$ and $0.040]$. 
A set of template spectra was generated for each one, and each model
galaxy was run through the continuum-subtraction software five times, once for
each metallicity. 
As demonstrated, quite significantly inaccurate results can be returned 
if the contrast between real and assumed metallicity is large (factors greater
than around 2).
However, using method {\sc v} and for all cases with weak \lya\ emission (\wlya=10\AA), 
the discrepancy in \lya\ fluxes is reduced to less than the returned statistical
error-bar
provided that $(a)$ metallicities are not discrepant by more than one adjacent 
step in the grid, and $(b)$ $S/N=5$ or greater has been obtained in all bands.
Fortunately, very strong metallicity gradients are not often found across
starburst regions and a single long-slit spectrum is likely to provide
sufficient information.

\subsubsection{The initial mass function}

The {\em lower} panel of Figure~\ref{fig:std_w10_metimf} shows the normalized returned
values of \wlya\ for the models with modified IMFs: the flatter 
{\tt imf185\_w10} and the steeper {\tt imf285\_w10}. 
IMFs are modified in both the starburst and underlying stellar components for
the generation of the template spectra. 
Methods {\sc iii} and {\sc iv} here appear to perform in a similar manner to when
metallicity is modified: normalized \wlya\ deviates further from unity with
increasing $S/N$ as poor fits are consistently found. 
However, the two-component stellar fit method always recovers a mean 
\wlya\ that scatters around the 1-line.
This results from the fact that IMF and star-formation history both alter the
current stellar mass distribution: fitting two components appears to mimic the
effect of a singly modified IMF.

\subsubsection{Stellar atmosphere}

Changing the UV stellar atmosphere model makes no appreciable difference to any
of the methodologies. 
All SED-fitting methods appear to return \wlya\ consistently around the desired
value and results are indistinguishable from those presented in
Figure~\ref{fig:std_w10_s2n}.
While testing one stellar atmosphere model against another in this manner may not 
be greatly meaningful, it is at least reassuring that the choice of atmosphere
does not impact upon the recovered values. 
The stellar atmosphere models differ quite significantly in their ionizing
output and in the detail of line features. 
However the ionizing output only affects the nebular fluxes which are treated 
independently, and the discrete line features have only a minor impact upon integrated 
colors, even in the FUV.

\section{Future studies and possible further improvements}\label{sect:future}

So far we have only discussed simulations relating to the data-sets we have already
obtained, targeting the lowest redshifts possible. 
However, this redshift severely limits the number of potential target galaxies,
and any future \lya\ imaging studies will require larger volumes, especially in
order to target the more luminous analogues of high-$z$ star-forming galaxies.
Beyond $z\approx 0.03$ the optimal choice of filters changes, and can
be performed using {\em ACS/SBC} and adjacent pairs of long-pass filters.
Here we present some simulations for such a study.
We also discuss some methodological alternatives using augmented datasets.

\subsection{A potential local \lya\ imaging study with HST}

As discussed in Sect.~\ref{sect:obsstrat} and illustrated in
Fig.~\ref{fig:filters}, we adopt the combination of {\em F125LP/F140LP},
restricting us to the broad redshift range of $0.028 - 0.09$, 
although any adjacent pair of long-pass filters can be used.
With the current inactive status of {\em ACS} CCD channels and uncertain future
of {\em WFC3}, we opt for 
bandpasses available on {\em WFPC2} to cover the optical domain, although this
setup could easily be ported to configurations on both these cameras.
Filters chosen for the continuum and \halpha\ observations are listed in 
Section.~\ref{sect:sedgen}
It is also worth noting that with this configuration, we have no medium-band 
line-free filter near \halpha\ and the continuum subtraction of \halpha\ must
rely on interpolation between {\em F439W} and {\em F814W}. 
Simulations presented here adopt the redshift $z=0.029$ so as to shift \lya\
into the on-line bandpass without shifting \halpha\ out of {\em F673N}.

\begin{figure*}
\epsscale{1} 
\plotone{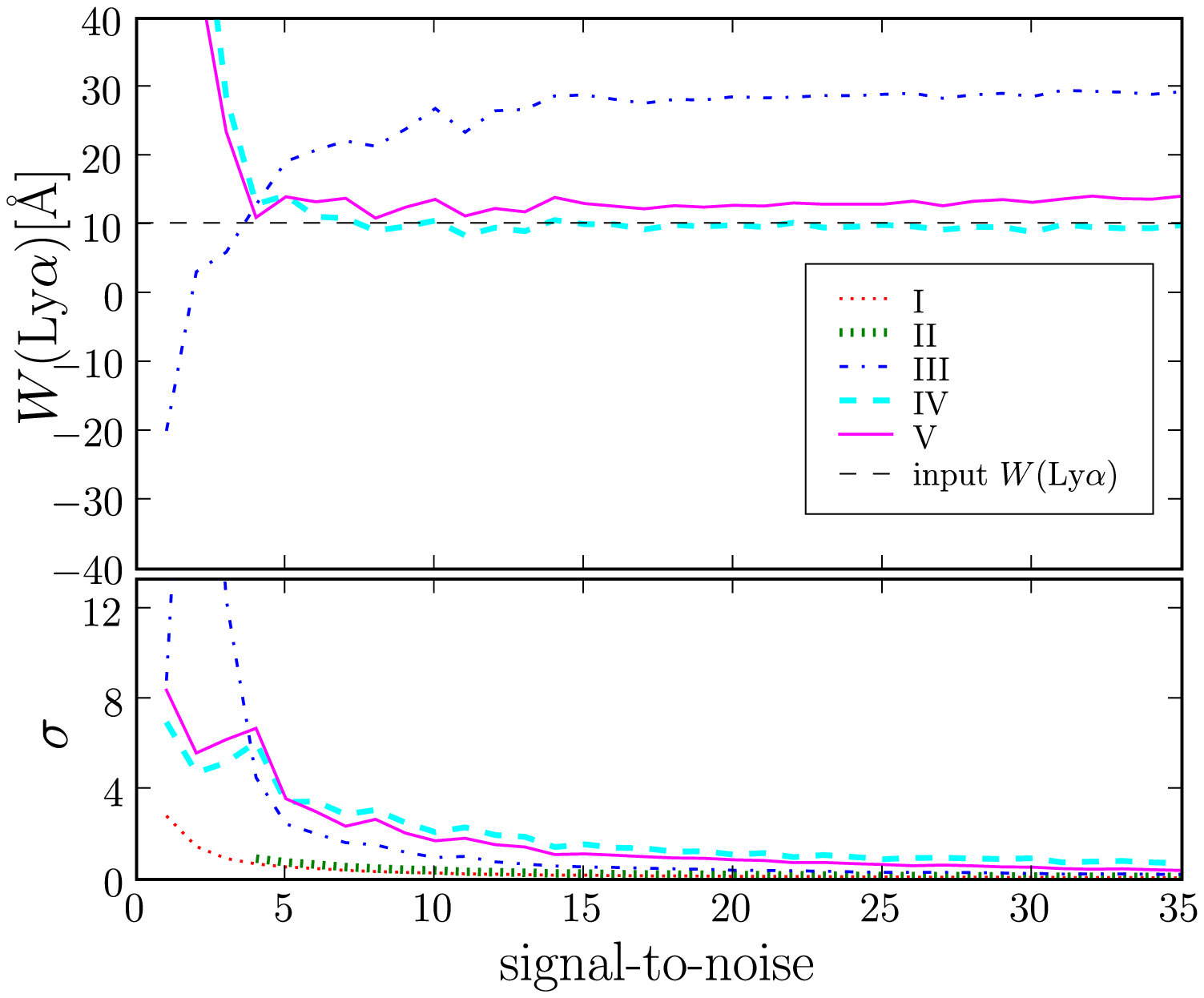}
\plotone{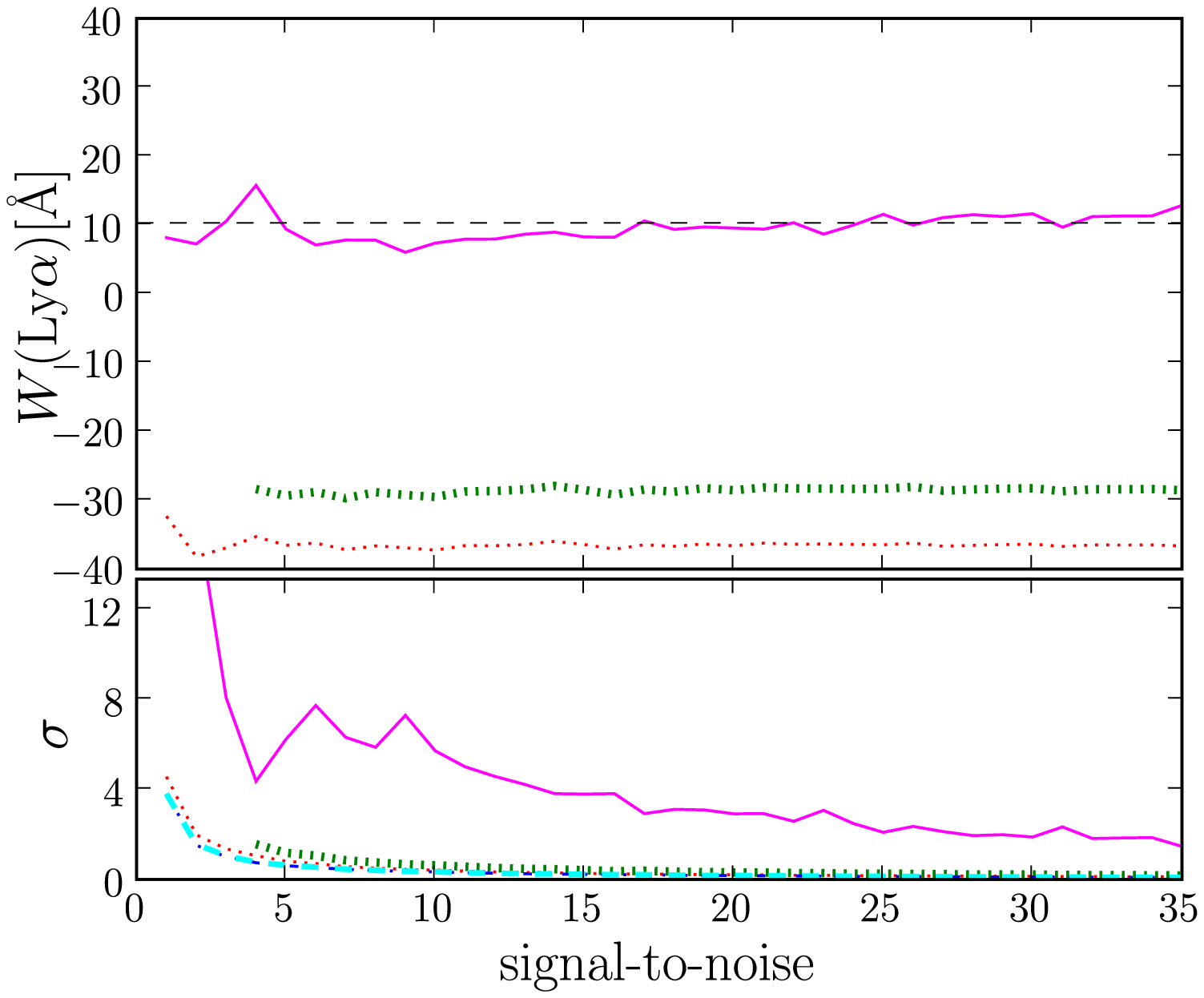}
\caption{Example results using observing configuration 2. Input \wlya\ is 10\AA. 
{\em Left}: nebular emission-dominated region ({\tt nneb1\_w10}). 
Again, methods {\sc i} and {\sc ii} converge well below the lower limit of the
ordinate. 
{\em Right}: region dominated by and old stellar population ({\tt nfs5\_w10}).
Methods {\sc iii} and {\sc iv} converge above the upper limit of the ordinate.
All continuum subtraction methods are shown and color-coded as in 
Figure~\ref{fig:std_w10_nneb_nfs}.
}
\label{fig:std_w10_lars}
\end{figure*}

Figure~\ref{fig:std_w10_lars} shows some example results from simulations
using configuration 2. 
They are the same results as shown for configuration 1 in
Figure~\ref{fig:std_w10_nneb_nfs}.
These Figures show the recovered \lya\ equivalent width from the 
nebular gas dominated and field-star dominated templates, and
are highly resemblant of those presented for configuration 1. 
Configuration 2 has the slight advantage over 1 of using two filters with
near-identical red wings, which isolates a well-defined on-line bandpass
shifted slightly nearer to the pivotal wavelength of the continuum
measurement.
However, configuration 2 does still suffer from the same drawback of a wide on-line
bandpass and, for $S/N=10$ in all bands, $S/N$ in the continuum subtraction is
still around 0.5 for the {\tt fid\_w10} model, the same as configuration 1
(Section~\ref{sect:goodsn}). 

Configuration 2 does include two fewer continuum filters than 1, which was
previously shown (Section~\ref{sect:poorsn} and 
Figure~\ref{fig:std_w10_rat_f220w_f330w}) to be inadequate: the lower panel 
shows how \wlya\ is very poorly recovered when the NUV and $U-$band filters 
are removed. 
This was designed to exemplify the poor recovery of \lya\ observables and this
case was especially bad due to the loss of both UV filters, and no sampling of
the FUV slope or Balmer/4000\AA\ break. 
The inclusion of both {\em F220W} and {\em F330W} for configuration 1 is likely 
to include some level of redundancy.
The continuum filters of configuration 2 are ideally placed and the most
important spectral features remain well sampled: 
{\em F140LP+F336W} samples $\beta$, 
{\em F336W+F439W} samples the Balmer/4000\AA\ break, 
and {\em F439W+F814W} simultaneously allows for the continuum subtraction of
\halpha\ and the constraint of the field star population. 
In light of these results and those presented in Section~\ref{sect:poorsn}, we
determine this to be the the absolute minimum requirement in SED coverage.

\subsection{Stellar and nebular extinction}

One area in the methodology that could be identified as a weakness is the
treatment of reddening: we treat stellar (\ebvs) and interstellar extinction (\ebvg)
in the same, possibly sub-optimal, way: by locking them together.
\ebvs\ and \ebvg\ are known to differ quite substantially in some cases, with
derived values of \ebvs\ lower than \ebvg\
\citep{Fanelli88,Calzetti94}. 
This seems to be a geometric effect due to winds from massive stars 
expelling the ionized ISM and dust, reducing the extinction
derived from the stellar UV slope and concentrating the dust in the \hii\
shells and filamentary structures
\citep{Maiz-Apellaniz98}.
\cite{Mas-Hesse_Kunth99}
found that the discrepancy grew with
increasing age over timescales consistent with stellar evolution and supernova 
enrichment. 
It is frequently the case that we can resolve nebular 
structure down to the resolution limit, although we will can never know the
geometry along the line--of--sight.
How the dust in front of stars and in ionized regions combine depends upon the 
unresolved ISM geometry and it may, in such cases, be preferable to treat
nebular and stellar reddenings independently in the method. 

Decoupling could be achieved simply in the fitting procedure, by introducing \ebvg\ 
as an extra fitting dimension that applies only to the nebular SED, allowing both
values of \ebv\ to vary independently. 
However, the number of SED data-points is not sufficient for the inclusion of an 
extra degree of freedom, and currently we deem this too computationally expensive to include 
in the fitting algorithm when we have $\gtrsim 10^6$ pixels per image.
While an empirical relationship has been presented between the two quantities by 
\cite{Calzetti00}: 
\ebvs~$=(0.44\pm 0.033)$~\ebvg, it was derived for a sample of galaxies observed
with much poorer resolution and, in individual {\em HST} pixels, the spread between the 
quantities is likely to be so large that the two quantities are completely decoupled.
Alternatively, \ebvg\ may be measured directly by using an emission line decrement. 
Typically the Balmer decrement \halpha/\hbeta\ would be used where \hbeta\ could
also be obtained from {\em HST} using linear ramp or selected narrowband
filters. 
A further possibility 
would be to use a NIR emission line (e.g. \paalpha\ or \brgamma) observed from 
the ground using a large adaptive optics imager.
Indeed, an independent evaluation of the nebular reddening is precisely what is
required for an appropriate astrophysical comparison with \lya.
A direct measurement of the interstellar reddening would permit us to lock
\ebvg\ on the nebular gas component and fit \ebvs\ to the stellar continuum.
To this end we are in the process of observing a subset of our current sample in
the \brgamma\ line. 
Such methods will be tested in a forthcoming study.

Ultimately, however, the continuum we need to estimate and subtract is predominantly 
stellar, and it may simply be that treatment of reddening on the stellar continuum is 
all that is required.
Further dedicated tests will be presented when the data have been acquired and 
processed.

\section{Conclusions and Summary}\label{sect:conc}

Using synthetic spectra of starburst galaxies we have examined various methods
of producing continuum subtracted line-only \lya\ images using currently available
imaging modes on the {\em Hubble Space Telescope}. 
We have assessed and compared various methods of continuum subtraction that vary in 
their complexity and attention paid to possible behavior of the continuum. 
We have presented examples covering a wide array of starburst
parameters for two observational configurations: 
our {\em ACS/SBC} imaging campaign of local starbursts 
(online/offline ={\em F122M / F140LP}), and for studies using adjacent pairs
of long-pass filters ({\em F125LP / F140LP}), targeting slightly higher-$z$.
Our main conclusions are: 
\begin{itemize}

\item{Making simple assumptions about the shape of the far ultraviolet
continuum slope ($\beta$; e.g. assuming its slope or extrapolating the slope from 
observations on the red side only) leads to estimates of the \lya\ flux and 
equivalent width that are seriously discrepant with the true values. Some 
spectral fitting is shown to be essential, even for the most basic cases.}

\item{In our methodology we fit only the age of the stellar 
population and the dust reddening, requiring at least datapoints on the SED that
sample the UV continuum slope and Balmer/4000\AA\ discontinuity. All other
parameters (extinction law, metallicity, IMF etc.) take assumed values; we then
investigate the systematic effects incurred when these quantities differ from
their true values.  }

\item{We need to bin together pixels until signal--to--noise of between 5 
		and 10 has been obtained.}

\item{Age determinations may be contaminated by boosted nebular gas emission
and an underlying stellar population.
These can be accounted for by constraining the nebular gas contribution using
an estimate based upon the \halpha\ emission flux, and contribution from old
stars by fitting multiple stellar components, respectively. 
This requires a data-point
redwards of \halpha, both for the continuum subtraction of \halpha\ and an
estimate of the contribution from field stars. 
We have determined the $I-$band to be functional.  }

\item{The initial mass function is shown not to be a necessary parameter to
include in the fitting process and results are largely unchanged when
non-standard IMFs are tested. Metallicity has a much more significant impact upon
the recovery of the \lya\ flux, since it changes the rate of stellar evolution
and reddens stellar continua. This requires an independent determination of
the metallicity. The choice of stellar atmosphere model has no
discernible impact upon the recovered \lya\ observables. }

\item{If the metallicity and reddening law are known quantities and $S/N=5$ has
been obtained in all bandpasses, no significant systematic effects are seen in our 
continuum subtracted fluxes or equivalent widths. 
Increasing $S/N$ to 10 is shown to significantly reduce the scatter. 
For input \wlya=10\AA, we are able to recover \lya\ fluxes 
accurate to within 30\% of the true value for all tested parameter space. 
This improves to better than 10\% for stronger \lya\ emission with \wlya=100\AA.}

\item{We have also presented simulations for a very similar study that uses
adjacent combinations of {\em SBC} long-pass filters to isolate \lya.
Due to the near-identical red wings, this could naively be thought to mitigate
many of the issues surrounding continuum subtraction. However, due to the braod
nature of the bandpass, we still determine a similar level of care to be 
necessary in the method.
}

\end{itemize}

\acknowledgments

MH and G{\"O} gratefully acknowledge the support of the Swedish National Space
Board (SNSB; Rymdstyrelsen) and Swedish Science Council (Vetenskapsr{\aa}det;
VR). 
JMMH is supported by Spanish MEC grant AYA2004-08260-C03-03.
We thank Hakim Atek for thoughtful comments on the manuscript, and Claus Leitherer
and Artashes Petrosian for their invaluable contribution to the ongoing
projects.

{\it Facilities:} \facility{HST (ACS)}.

\end{document}